\documentclass[aps,prl,twocolumn,reprint]{revtex4-2}
\usepackage[utf8]{inputenc}
\usepackage{bm}
\usepackage{amssymb}
\usepackage{mathtools}
\usepackage{amsmath}
\usepackage{xcolor}
\usepackage{enumitem}
\usepackage{natbib}
\usepackage{hyperref}
\usepackage{graphicx}

 \hypersetup{colorlinks=true, linkcolor=blue, breaklinks=true, urlcolor=blue, citecolor=blue}
\usepackage{lipsum}
\usepackage[ruled,vlined]{algorithm2e}

\begin{document}

\title{Predicting the response of structurally altered and asymmetrical networks} 
\author{Melvyn Tyloo}
\affiliation{Living Systems Institute, University of Exeter, Exeter, EX4 4QD, United Kingdom\\ Department of Mathematics and Statistics, Faculty of Environment, Science, and Economy, University of Exeter, Exeter, EX4 4QD,
United Kingdom}
\date{\today}

\begin{abstract}
We investigate how the response of coupled dynamical systems is modified due to a structural alteration of the interaction.
The majority of the literature focuses on additive perturbations and symmetrical interaction networks.  
Here, we consider the challenging problem of multiplicative perturbations and asymmetrical interaction coupling. 
We introduce a framework to approximate the averaged response at each network node for general structural perturbations, including non-normal and asymmetrical ones.
Our findings indicate that both the asymmetry and non-normality of the structural perturbation impact the global and local responses at different orders in time. 
We propose a set of matrices to identify the nodes whose response is affected the most by the structural alteration. 
\end{abstract}

\maketitle
\textit{Introduction}.-- Interaction among dynamical systems leads to various collective emergent behaviors ranging from spontaneous magnetization in materials to opinion polarization on social networks~\cite{yang1952spontaneous,vicsek1995novel,Pec98, baumann2020modeling}.
The structure of the interaction coupling, including its strength essentially determines the response of such macroscopic states to external perturbations. 
The quantification of the response of coupled dynamical systems is an important open problem that keeps drawing the attention of physicists and mathematicians~\cite{Men13,ronellenfitsch2018optimal,Tyl18a}. 
Typically these systems are modelled using networks where the nodes and edges represent, respectively, the dynamical units and interaction coupling~\cite{newman2018networks,posfai2016network}.
One then evaluates the network properties and correlates them with particular features of the response~\cite{pastor2001epidemic,baumann2020periodic,tyloo2022layered}. 

A challenging question is to evaluate how the response is altered when the dynamics itself is modified, i.e. the network structure through the edges is modified.
Especially when the interaction becomes asymmetrical and non-normal, many mathematical relations cannot by applied anymore, leading to fewer analytical predictions. 
However, many real-world networked systems display asymmetry and non-normality~\cite{asllani2018structure}.
Any effort in answering this question might have direct application in the control of systems in fields as diverse as active flow and information networks, electric power grids including losses, neuron dynamics, network inference, non-Hermitian quantum mechanics~\cite{gao2016universal,ashida2020non,ronellenfitsch2021optimal,liu2022network,tyloo2023assessing,succar2025detecting}. 

In this letter, we focus on coupled dynamical systems whose dynamics can be approximated by their linear dynamics around a stable fixed point. 
The structural alteration of the interaction is modelled as follows. 
Starting with symmetrical coupling interaction, we introduce a multiplicative perturbation that effectively makes the interaction asymmetrical in general. 
Borrowing tools from quantum mechanics to approximate the product of two matrix exponentials, we provide an analytical estimation of the amplitude of the response of each dynamical unit.
We show that the short time response is given by functions of three matrices involving the dynamical matrix and the structural perturbation. 
Their diagonal elements can be used to identify the nodes that are the most affected by the structural perturbation. 
Interestingly, we show that the global averaged response is affected by the asymmetrical part of the structural perturbation later than the local response at each node. 
Eventually, we show that one can use our framework to assess the response of unaltered asymmetrical network.

\textit{Theoretical framework}.-- We focus on the linear coupled dynamical system,
\begin{align}\label{eq1}
    \dot{\bf x} = ({\bf A} + {\bf \Delta}){\bf x}  + {\bm \eta}
\end{align}
where ${\bf x}\in\mathbb{R}^N$ are the degrees of freedom at the $N$ nodes and ${\bf A},{\bf \Delta}\in\mathbb{R}^{N\times N}$ are, respectively, the dynamical matrix and the structural perturbation. 
They define the interaction coupling between the network nodes. 
The matrix ${\bf A}$ is symmetric, which corresponds to the initial symmetrical coupling, i.e ${\bf A}={\bf A}^\top$\,, while ${\bf \Delta}$ is an asymmetrical matrix that models the structural alteration of the interaction, i.e. ${\bf \Delta}\neq {\bf \Delta}^\top$\,. 
We additionally require that all the eigenvalues of ${\bf A}$ are non-positive. 
Note that self-loops could also be included. 
The last term $\bm \eta$ models any external additive perturbation. 
The initial condition is taken to be ${\bf x}(t=0) = {\bf 0}$\,.
It is important to remark that Eq.~(\ref{eq1}) can be viewed as the linearization of a nonlinear dynamical system around a stable fixed, in which case, $({\bf A} + {\bf \Delta})$ would correspond to the Jacobian of the system~\cite{strogatz2024nonlinear}. 

We are interested in assessing how the response of the system is modified by the structural perturbation $\bf \Delta$\,. 
The response to a specific choice of $\bf\eta$ is not what we are aiming for, but rather the response averaged over an ensemble of perturbations spanning all the potential input nodes. 
Let us take as an ensemble of perturbations ${\bm \eta}(i,t) = \hat{\bf e}_i\,\delta(t)$ with $i=1,...,N$\,, where $\hat{\bf e}_i$ is the $i$-th canonical vector. 
This choice of ensemble will have useful properties in later derivations. 
Such ensemble are also used in control theory to evaluate performance metrics~\cite{Zhou1998Essentials,jovanovic2008h2,poolla2017optimal,paganini201755th}.
With the vanishing initial condition, one obtains the response to the perturbation ${\bm \eta}(i,t)$\,,
\begin{align}
    {\bf x}(i,t) = [e^{({\bf A}+{\bf \Delta})t}]_{.,i}\,,
\end{align}
where $[C]_{.,i}$ denotes the $i$-th column of the matrix $C$\,. 
To quantify the amplitude of the response following the perturbation, we focus on the square of the trajectories at each node. 
As stated earlier, we are not interested in a specific perturbation but rather the expected response to any perturbation. 
 We therefore take the average over the ensemble of perturbations ${\bm \eta}(i,t)$\,, $i=1,...N$\,. We denote the averaged squared response at node $j$ as $\overline{x_j^2}(t)=\sum_{i=1}^N[{\bf x}(i,t){\bf x}(i,t)^\top]_{jj}$\,. Doing so, one has,
\begin{align}\label{eq:res}
    \overline{x_j^2}(t)= [e^{({\bf A}+{\bf \Delta})t}e^{({\bf A}^\top+{\bf \Delta}^\top)t}]_{jj}\,,
\end{align}
The above equation gives the time-evolution of the averaged amplitude of the response at node $j$\,. 
Of course, assuming that the matrix $({\bf A}+{\bf \Delta})$ is non-defective, one can diagonalize it and obtain an expression in terms of its left and right eigenvectors and corresponding eigenvalues. 
If instead the matrix is defective, one cannot perform the eigen-decomposition. 
Either way, one then relies on numerical investigation to assess the effect of $\bf\Delta$ on the response. 
Here instead, we want to deepen our analytical understanding about how the asymmetry modifies the response of the system. 
So let us forget about performing any eigen-decomposition this time. The reason behind it will become clear below.

\begin{figure}
    \centering
    \includegraphics[width=1.0\linewidth]{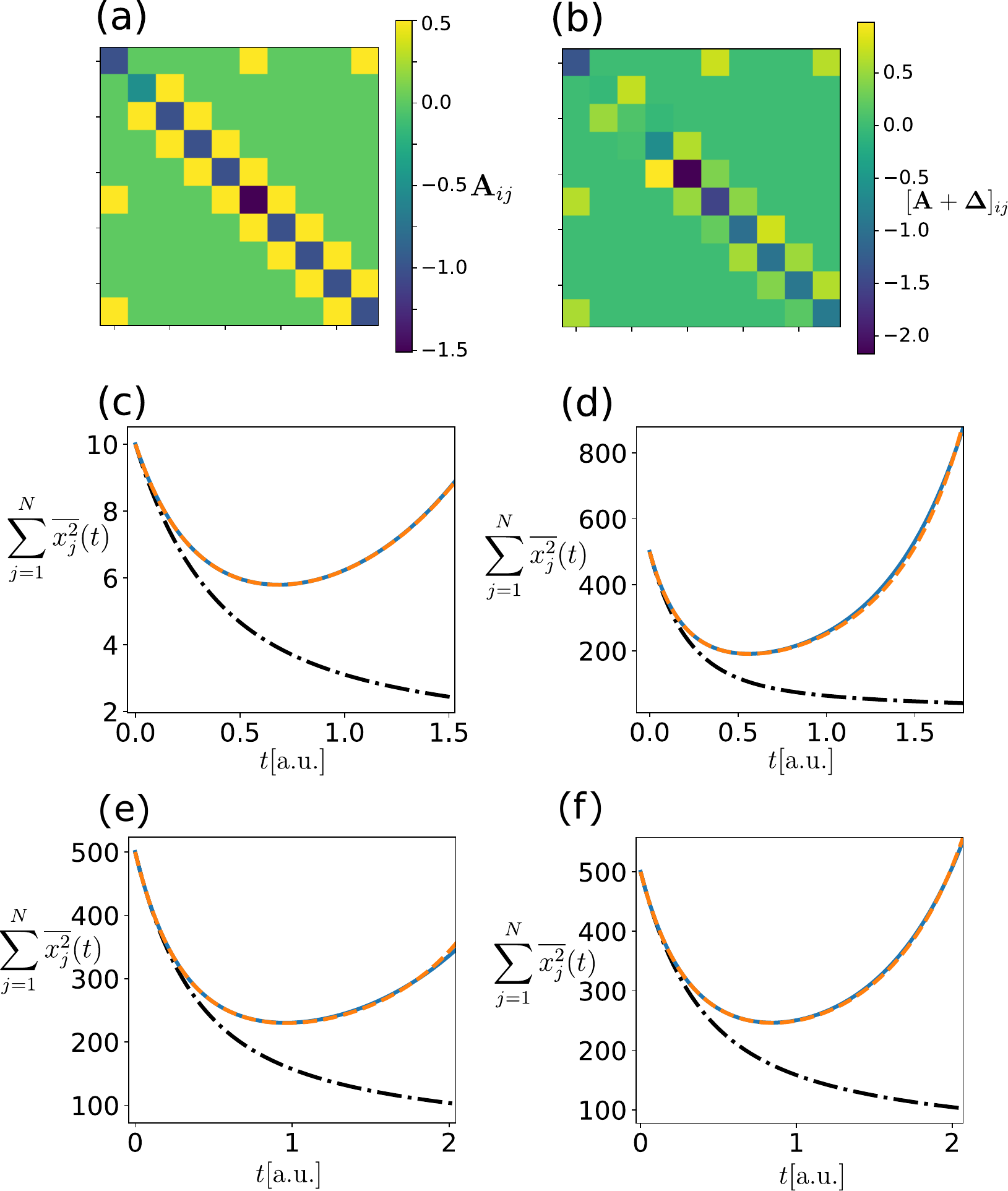}
    \caption{
    Comparison between theory Eq.~(\ref{eq:tr}) and numerical time-evolution of Eq.~(\ref{eq1})\,.
    (a) Matrix elements of $\bf A$ used for panel (b) and (c).
    It is the negative of the Laplacian matrix of a Watts-Strogatz network~\cite{watts1998collective} with $N=10$ nodes, $m=2$ initial nearest-neighbors and edge rewiring probability $p=0.2$\,, to which self loops of weight $0.01$ were added. 
    (b) Matrix elements of $({\bf A} + {\bm \Delta})$\,. 
    The perturbation matrix ${\bm\Delta}$ is given by ${\bm\Delta}_{ij}={\bf A}_{ij}\xi_{ij}$ with $\xi_{ij}\sim \mathcal{N}(0,0.25)$ and $i,j=1,...N$\,. 
    (c) Sum of the square of the trajectories averaged over the ensemble of perturbations ${\bm \eta}(i,t) = \hat{\bf e}_i\,\delta(t)$ with $i=1,...,N$\,. 
    The blue solid line is obtained by numerically solving Eq.~(\ref{eq1})\,. 
    The orange dashed line gives the right-hand side of Eq.~(\ref{eq:tr})\,. 
    The dotted-dashed line is obtained by numerically solving Eq.~(\ref{eq1}) without the structural perturbation matrix, i.e. setting ${\bf \Delta}=\bf 0$\,.
    (d), (e), (f) Same plots as (c) but using three different types of network with $N=500$ nodes. 
    (d) Watts-Strogatz network with $m=4$ initial nearest-neighbors and edge rewiring probability $p=0.05$\,. 
    (e) Barab\'asi-Albert network with $m=4$ new neighbors~\cite{newman2018networks}.
    (f) Erd\H{o}s-R\'enyi network with edge probability $p=0.05$\,~\cite{newman2018networks}. 
    For all three networks self loops of weight $0.01$ were added.
    }\label{fig1}
\end{figure}
In the most general case, one has the commutator $[{\bf A} + {\bf \Delta},{\bf A}^\top + {\bf \Delta}^\top]\neq 0$\,, namely $({\bf A}+{\bf \Delta})$ is a non-normal matrix.
This has important consequences as one cannot simply add up the exponents in Eq.~(\ref{eq:res})\, to obtain a single matrix exponential. 
Instead, using the Baker–Campbell–Hausdorff formula, Eq.~(\ref{eq:res}) can be written as
\begin{align}\label{eq:res2}
    \overline{x_j^2}(t) &= [e^{{\bf M}_1\,t + {\bf M_2}\,\frac{t^2}{2} + {\bf M_3}\,\frac{t^3}{12}+h.o.t.}]_{jj}\,,
\end{align}
where we defined the matrices 
\begin{align}
{\bf M}_1 &= 2{\bf A}+{\bf \Delta}+{\bf \Delta}^\top\,,\label{eq:M1}\\
{\bf M}_2 &= [{\bf A}, {\bf \Delta}^\top- {\bf \Delta}] + [{\bf \Delta},{\bf \Delta}^\top]\,,\label{M2}\\
{\bf M}_3 &= [{\bf \Delta}- {\bf \Delta}^\top, [{\bf A}, {\bf \Delta}^\top- {\bf \Delta}] + [{\bf \Delta},{\bf \Delta}^\top]\,]\,.\label{eq:M3}
\end{align}
We also used $[{\bf A} +{\bf \Delta},{\bf A}^\top + {\bf \Delta}^\top]=[{\bf A}, {\bf \Delta}^\top-{\bf \Delta}] + [{\bf \Delta},{\bf \Delta}^\top]\,,$\, and omitted the higher order terms beyond $t^3$ in the exponent of Eq.~(\ref{eq:res2})\,. 
One remarks that both matrices ${\bf M}_1$ and ${\bf M}_2$ are symmetric while ${\bf M}_3$ is skew-symmetric. As a sanity check, one indeed recovers simply the sum of the exponents in Eq.~(\ref{eq:res}) when $({\bf A}+{\bf \Delta})$ is normal, as only ${\bf M}_1$ remains non-vanishing.
From Eq.~(\ref{eq:res2}), one notices that as time grows, different products involving matrices ${\bf A}$\,, ${\bm\Delta}$ and ${\bm \Delta}^\top$ contribute to the averaged response. 
Indeed, at short times right after the perturbation, the dominating term is given by the symmetric matrix $(2{\bf A}+{\bf \Delta}+{\bf \Delta}^\top)$\,. 
Interestingly, a consequence of this is that the leading order in $t$ of the short time averaged response is independent of the asymmetry introduced by ${\bm \Delta}$\,. 
Therefore, different choices of ${\bm \Delta}$ with the same symmetric matrix $({\bm \Delta}+{\bm \Delta}^\top)$ yield the same short time response. 
Moreover, if ${\bf \Delta}$ is skew-symmetric, then there is no linear correction in $t$ induced by the structural perturbation. 
Both the asymmetry of ${\bm \Delta}$ through the commutator $[{\bf A}, {\bf \Delta}^\top- {\bf \Delta}]$\,, and its non-normality with $[{\bf \Delta},{\bf \Delta}^\top]$ come into play at the next leading order in $t$ in the exponent of Eq.~(\ref{eq:res2})\,. 
One thus expects to observe the effect of the asymmetry of ${\bm \Delta}$ later in the response. 
Eq.~(\ref{eq:res2}) already provides important information about when the properties of ${\bm \Delta}$ impact the response. 
Also, provided that $\bf \Delta$ is not too large, one may approximate Eq.~(\ref{eq:res2}) as,
\begin{align}\label{eq:res3}
    \overline{x_j^2}(t) &\cong [e^{{\bf M}_1\,t + {\bf M_2}\,\frac{t^2}{2} + {\bf M_3}\,\frac{t^3}{12}}]_{jj}\,,
\end{align}
where we dropped the h.o.t in the exponent. 
Below we check the validity of the approximation. 
Before that, let us gain more insights by focussing first on the global response of the network and then move on to the local response of each node.

Investigating the global response at the network level might inform us on how to pick a specific matrix ${\bm \Delta}$ that would simultaneously modify the response at the nodes the most. 
As a measure of the global response of the network, one can take the sum of Eq.~(\ref{eq:res3}) over all the nodes, which can be written with the trace,
\begin{align}\label{eq:tr}
    \sum_{j=1}^N\overline{x_j^2}(t) \cong {\rm Tr}\left[e^{{\bf M}_1\,t + {\bf M_2}\,\frac{t^2}{2} + {\bf M_3}\,\frac{t^3}{12}}\right]\,.
\end{align}
  \begin{figure}
    \centering
    \includegraphics[width=0.8\linewidth]{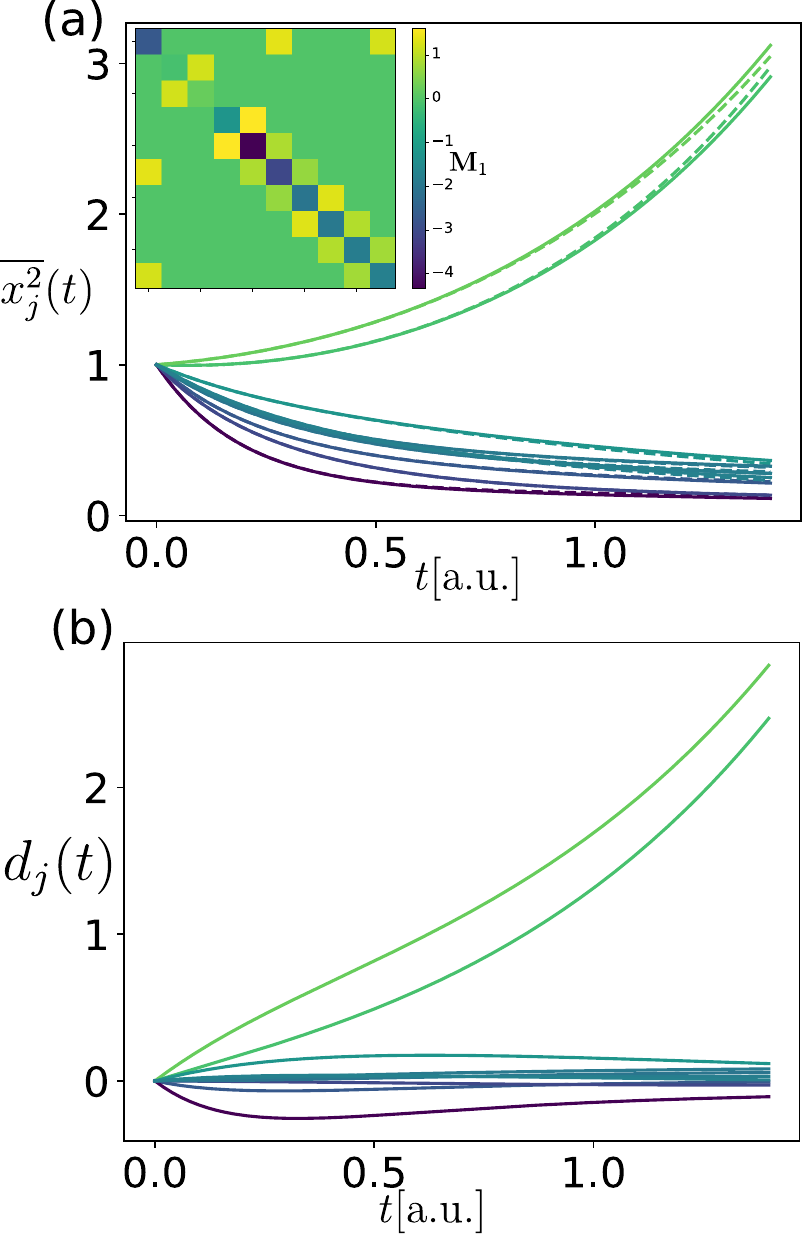}
    \caption{
     Comparison between theory Eq.~(\ref{eq:res3}) and numerical time-evolution of Eq.~(\ref{eq1})\,.
    (a) Square of the averaged response at each node of the network. 
    The matrix $\bf A$ and the structural perturbation $\bf\Delta$ are the same as in Fig.~\ref{fig1}(c)\,.
    Numerical simulations are given by the solid lines. 
    The predictions of Eq.~(\ref{eq:res3}) are given by the dashed lines. 
    Inset: matrix elements of ${\bf M}_1$ [see Eq.~(\ref{eq:M1})]\,.
    (b) Difference between the response without and with the structural perturbation [$d_j(t) = \overline{x^2_j}({\bf A}+{\bf\Delta},t)-\overline{x^2_j}({\bf A},t)$]\,. 
    In both panels, the color of the trajectories are given the diagonal of the matrix ${\bf M}_1$ depicted in the inset of (a)\,.
    }\label{fig2}
\end{figure}
Many interesting observations have to be made about Eq.~(\ref{eq:tr})\,, but let us start by checking whether it is a fair approximation of the response. 
In Fig.~\ref{fig1}, we numerically simulate Eq.~(\ref{eq1}) with ${\bf A}=-{\bf L}(G)$ where ${\bf L}(G)$ is the Laplacian matrix of a Watts-Strogatz network~\cite{watts1998collective} [Fig.~\ref{fig1}(a)] with $N=10$ nodes. 
The structural perturbation $\bf\Delta$ is randomly drawn (see caption) and its effect is given in Fig.~\ref{fig1}(b)\,, which depict the matrix elements of $({\bf A} + {\bf\Delta})$\,.
For this network and perturbation, one observes in Fig.~\ref{fig1}(c) a good agreement between Eq.~(\ref{eq:tr}) (dashed orange line) and the simulation (blue solid line). 
One should also remark that $\bf\Delta$ significantly modifies the response of the network compared to the case without any structural perturbation (dotted-dashed line). 
In Fig.~\ref{fig1}(d), (e), (f) we further check the theory against numerical simulations for larger networks of $N=500$ nodes obtained using different network generating algorithms.
Note that the response of the network does not always diverge or is slower to return to zero. 
It can be decaying faster than the unperturbed network as we illustrate in the Supplemental Material~\cite{SM}.

This being checked, let us move back to Eq.~(\ref{eq:tr})\,. 
The first thing to notice is the correction to the short time behavior that is dictated by the symmetric part of the perturbation matrix i.e. $({\bf\Delta}+{\bf\Delta}^\top)$\,.
Thus, the short time response is mostly independent of the asymmetry of $\bf\Delta$\,, which only kicks in at higher order in $t$ in Eq.~(\ref{eq:tr})\,.
 We verify numerically this fact in the Supplemental Material~\cite{SM} where make use of many different perturbation matrices $\bf\Delta$ with the same symmetric part (${\bf\Delta}+{\bf\Delta}^\top$) but different asymmetric parts (${\bf\Delta}-{\bf\Delta}^\top$)\,.
 Second, it seems that at the next order in time, we find two contributions, namely the commutator between $\bf A$ and $({\bf\Delta}^\top-{\bf\Delta})$ and the commutator between $\bf\Delta$ and ${\bf\Delta}^\top$\, [see Eq.~(\ref{M2})]. 
 While the latter is directly related to the normality of $\bf\Delta$\,, the former term is more complex as it involves some interaction between $\bf A$ and the perturbation $\bf\Delta$\,. 
 Surprisingly, by scrutinizing Eq.~(\ref{eq:tr}), one actually remarks that both terms only affect the response at the fourth order in $t$\,. 
 Indeed, rewriting the first few terms of the exponential series of Eq.~(\ref{eq:tr})\,, one has,
  \begin{align}\label{eq:tr_long}
     \sum_{j=1}^N\overline{x_j^2}(t) &\cong N + {\rm Tr}[{\bf M}_1]\,t + {\rm Tr}[{\bf M}_1^2]\,\frac{t^2}{2} +{\rm Tr}[{\bf M}_1^3]\,\frac{t^3}{6} \nonumber\\
     &\hspace{-1cm} + \left({\rm Tr}[{\bf M}_1^4] + 2\,{\rm Tr}[{\bf M}_1{\bf M}_3] + 3\,{\rm Tr}[{\bf M}_2^2]\right) \, \frac{t^4}{24} + \mathcal{O}(t^5)
 \end{align}
 where we used that the trace of a commutator is always vanishing no matter what are the matrices, i.e. ${\rm Tr}\left([{\bf A},{\bf\Delta}^\top-{\bf\Delta}]\right)= {\rm Tr}\left([{\bf\Delta}^\top,{\bf\Delta}]\right) =0$.
  We also used the property of the anti-commutator that ${\rm Tr}\left[\{{\bf B},{\bf C}\}\right]=2{\rm Tr}[{\bf B}{\bf C}]$\,. 
  The first line of Eq.~(\ref{eq:tr_long}) only involves $\bf A$ and the symmetric part of the perturbation $({\bf\Delta}+{\bf\Delta^\top})$\,. 
  The asymmetrical part as well as the non-normality of the perturbation show up at the second line of Eq.~(\ref{eq:tr_long})\,, at the fourth order in $t$\,~\cite{rem1}. 
  More specifically, distinct structural perturbations ${\bf \Delta}$ that are skew-symmetric only start to differ from each other, and from the unperturbed response, at the fourth order in time, with the trace of the matrix $(2\, {\bf M}_1{\bf M}_3 + 3\,{\bf M}_2^2)$\,.
  Therefore, the asymmetry and non-normality of the structural perturbation $\bf \Delta$ only affect the averaged global response later in the transient. 

  While the latter results is true for the global averaged response of the network, interestingly it is not the same when looking at the local averaged response. 
  Before elaborating on that point, let us numerically check the approximation of Eq.~(\ref{eq:res3}) for the local response. 
  In Fig.~\ref{fig2}(a)\,, we show the response at each node for the same network and structural perturbation as in Fig.~\ref{fig1}(c)\,. 
   One observes a good agreement between the theory Eq.~(\ref{eq:res3}) (dashed lines) and the numerical simulations (solid lines)\,.
  The difference $d_j(t)$ between the trajectories at each node with the structural perturbation and the trajectories without the perturbation is shown in Fig.~\ref{fig2}(b)\,.
 One notices that the two nodes with the largest difference are identified by the diagonal of ${\bf M}_1$\, [see inset of Fig.~\ref{fig2}(a)]. 
  
\begin{figure}
    \centering
    \includegraphics[width=1\linewidth]{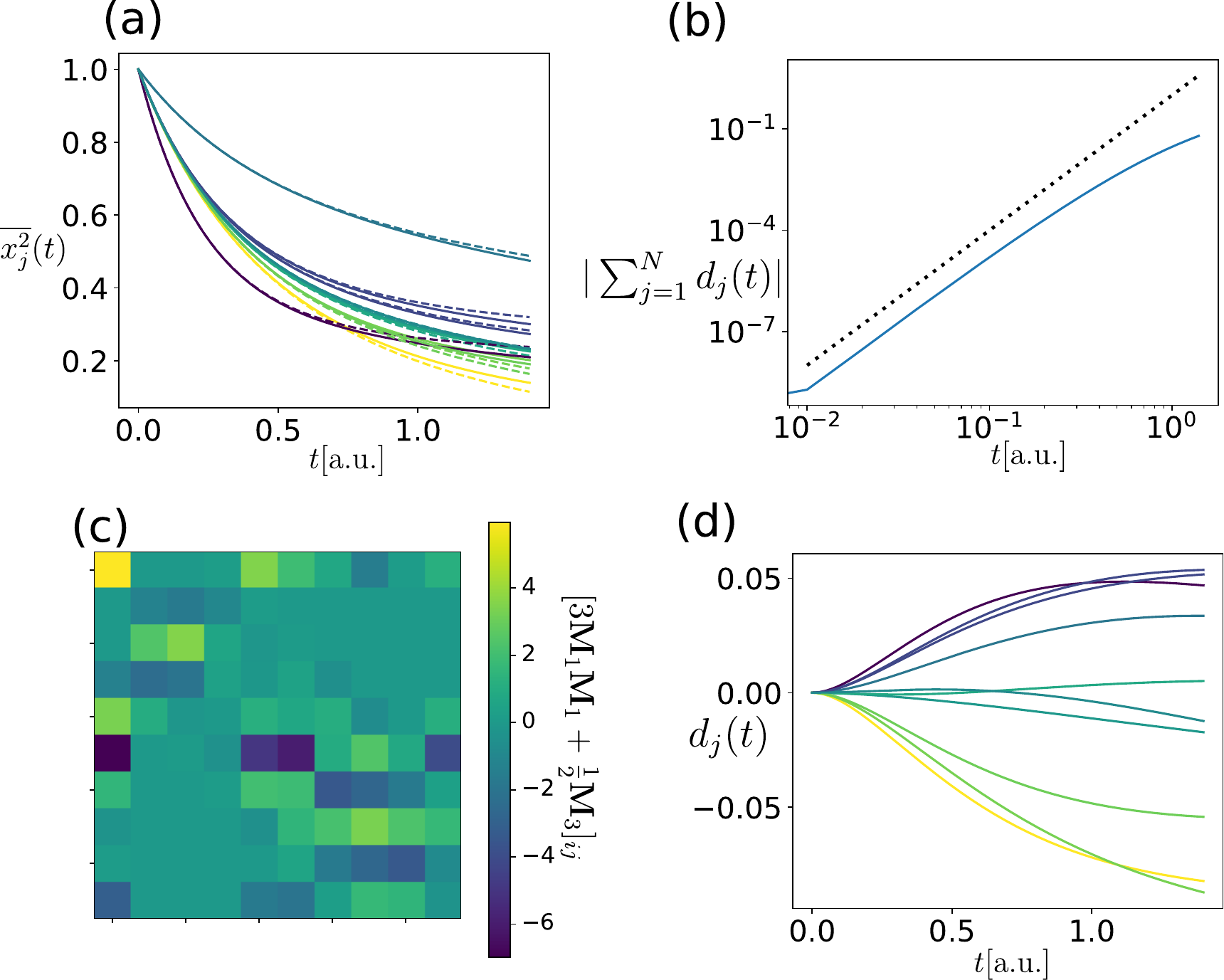}
    \caption{
      Comparison between theory Eq.~(\ref{eq:res3}) and numerical time-evolution of Eq.~(\ref{eq1})\,.
    (a) Square of the averaged response at each node of the network. 
    The matrix $\bf A$ is the same as in Fig.~\ref{fig1}(c)\,. 
    The structural perturbation $\bf\Delta$ is skew-symmetric and randomly obtained as ${\bf \Delta} = \frac{1}{2}(\overline{\bm\Delta}-\overline{\bm\Delta}^\top)$ where $\overline{\bm\Delta}_{ij}={\bf A}_{ij}\xi_{ij}$ with $\xi_{ij}\sim \mathcal{N}(0,0.25)$ and $i,j=1,...N$\,. 
    Numerical simulations are given by the solid lines. 
    The predictions of Eq.~(\ref{eq:res3}) are given by the dashed lines. 
    (b) Absolute value of the difference between the global response without and with the structural perturbation [$|\sum_{j=1}^N d_j(t) |= |\sum_{j=1}^N\overline{x^2_j}({\bf A}+{\bf\Delta},t)-\overline{x^2_j}({\bf A},t)|$]\,. 
    The dotted line give the scaling $\sim t^4$\,.
    (c) Matrix element of $(3{\bf M}_1{\bf M}_2 +  \frac{1}{2}{\bf M}_3)$\,.
    The trajectories of each node in panels (a) and (d) are colored according to the diagonal element of this matrix. 
    (d) Difference between the response without and with the structural perturbation [$d_j(t) = \overline{x^2_j}({\bf A}+{\bf\Delta},t)-\overline{x^2_j}({\bf A},t)$]\,. 
    }    
    \label{fig3}
\end{figure}

  We can now come back to the surprising result. Let us write down the first terms of the exponential series in Eq.~(\ref{eq:res3}),
  \begin{align}\label{eq:lo_long}
     \overline{x_j^2}(t) &\cong 1 + [{\bf M}_1]_{jj}\,t + \left([{\bf M}_1^2]_{jj} + [{\bf M}_2]_{jj}\right)\,\frac{t^2}{2} \nonumber \\
     &\hspace{-1cm}+  \left([{\bf M}_1^3]_{jj} +  3[{\bf M}_1{\bf M}_2]_{jj} +  \frac{1}{2}[{\bf M}_3]_{jj}\right)\,\frac{t^3}{6} + \mathcal{O}(t^4) 
  \end{align}
  Again, a few comments should be made about this expression. 
  First, unlike the global averaged response in Eq.~(\ref{eq:tr_long})\,, one has that the asymmetrical part and the non-normality of $\bf \Delta$ already affect the response at second order in $t$\,. 
  Second, if the structural perturbation is skew-symmetric i.e. ${\bf \Delta}^\top = - {\bf \Delta}$\,, the nodes whose response is mostly affected can be identified by the diagonal elements of matrices ${\bf M}_2$ and $(3{\bf M}_1{\bf M}_2 +  \frac{1}{2}{\bf M}_3)$\,.
  We illustrate numerically both these observations in Fig.~\ref{fig3}. 
  We take the same $N=10$ node network as in Fig.~\ref{fig1}(a) and choose a skew-symmetric structural perturbation $\bf\Delta$\,. 
  In Fig.~\ref{fig3}(a) one sees that the theory (dashed lines) and the numerical simulations (solid lines) agree well. 
  We check in Fig.~\ref{fig3}(b) that the leading order correction of the global response for a skew-symmetric perturbation is the fourth order in $t$\,, as predicted by Eq.~(\ref{eq:tr_long})\,. 
  Then, using the expression for the local averaged response in Eq.~(\ref{eq:lo_long})\,, one can identify the nodes whose response is the most affected by the structural perturbation. 
  For the realization of $\bf\Delta$ used in Fig.~\ref{fig3}\,, we found that the matrix $(3{\bf M}_1{\bf M}_2 +  \frac{1}{2}{\bf M}_3)$ has larger elements than ${\bf M}_2$\,. 
  Thus, the nodes that are the most affected are identified by the diagonal element of the former matrix, as shown in Fig.~\ref{fig3}(d)\,. 
  Comparing panels (b) and (d)  of Fig.~\ref{fig3}\,, one observes that even though the global response is barely altered by $\bf\Delta$ at short times, the local response is actually impacted by the structural perturbation.
  Finally, one should remark that the system in Fig.~\ref{fig3} can be viewed as an unaltered asymmetrical network. 
  Our framework thus allows to analyze the response of such systems as well.

\textit{Conclusions and outlook}.-- Predicting the response of coupled dynamical systems has broad implications in fields such as climatology, material science, social dynamics, neuroscience to name but a few. 
Specifically, understanding how the response is altered when the coupling interaction changes is of paramount importance to anticipate a potential loss of stability, or identify vulnerable network components. 

In this letter, we introduced a framework to approximate the averaged response of networks whose structure has been modified by a perturbation that can be non-normal and asymmetrical. 
We focussed on predicting the square of the trajectories at each node, and showed that the global averaged response is mostly independent of the asymmetry and non-normality of the structural perturbation at short times.
However, for the local averaged response, both the asymmetry and non-normality have effect at earlier times. 
Using the leading order terms of the matrix exponential series, we introduced a set of matrices Eqs.~(\ref{eq:M1})-(\ref{eq:M3}) to identify the nodes whose response is the most affected by the structural perturbation. 
They are obtained using the Baker-Campbell-Hausdorff formula, as combinations of the dynamical matrix and structural perturbation.
Note that our framework allows to analyze the response of structurally unperturbed asymmetrical networks.

The framework proposed in this letter opens the way to many interesting questions. 
We focussed on the response averaged over an ensemble of perturbations. 
It would be of interest to obtain an analytical prediction for the response to a specific perturbation, as some of them might be more likely to happen than others. 
Also, one should consider inputs other than Dirac-$\delta$ distributions, to better understand the interplay between time-scales within the system. 
Eventually, instead of using network generating algorithms, one should investigate real-world networks and analyze the matrices introduced here to identify the nodes that are the most altered in specific ways.

\pagebreak
\renewcommand{\thefigure}{S\arabic{figure}}
\setcounter{figure}{0}    
\onecolumngrid
\section{Predicting the response of structurally altered and asymmetrical networks: Supplemental Material}

\subsection{Comparison between theory and numerics: Faster decay}
\begin{figure}[h]
    \includegraphics[scale=0.45]{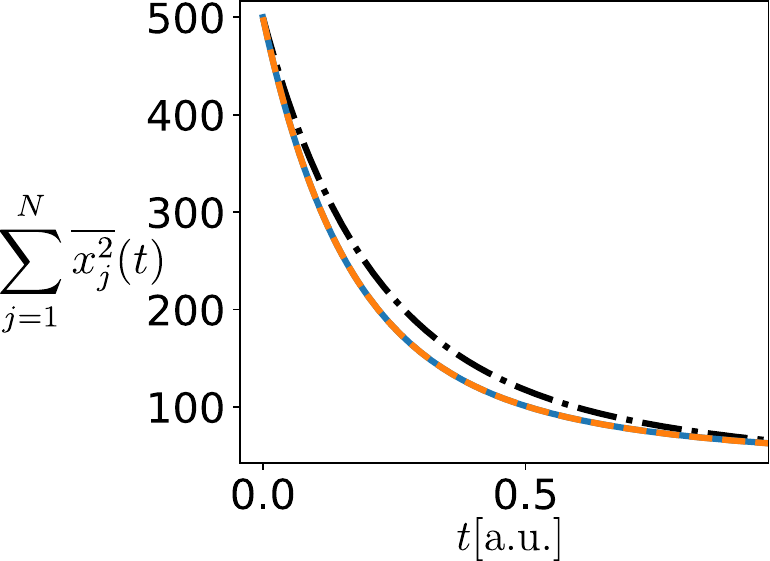}
    \caption{
    Comparison between theory Eq.~(\ref{eq:tr}) and numerical time-evolution of Eq.~(\ref{eq1}) in the main text.
    Sum of the square of the trajectories averaged over the ensemble of perturbations ${\bm \eta}(i,t) = \hat{\bf e}_i\,\delta(t)$ with $i=1,...,N$\,. 
    The matrix $\bf A$ is the same as in Fig.~\ref{fig1}(d) of the main text. 
    The perturbation matrix ${\bf\Delta}$ is given by ${\bm\Delta}_{ij}={\bf A}_{ij}\xi_{ij}$ with $\xi_{ij}\sim {\rm Unif}(0,0.5)$ and $i,j=1,...N$\,. 
    The blue solid line is obtained by numerically solving Eq.~(\ref{eq1})\,. 
    The orange dashed line gives the right-hand side of Eq.~(\ref{eq:tr})\,. 
    The dotted-dashed line is obtained by numerically solving Eq.~(\ref{eq1}) without the structural perturbation matrix, i.e. setting ${\bf \Delta}=\bf 0$\,.
    }\label{figS1}
\end{figure}

In Fig.~\ref{figS1}, we take the same Watts-Strogatz network as in Fig.~\ref{fig1}(d) but with a different structural perturbation which is now given by ${\bm\Delta}_{ij}={\bf A}_{ij}\xi_{ij}$ with $\xi_{ij}\sim {\rm Unif}(0,0.5)$ and $i,j=1,...N$\,.
In this case, one obtains a faster decay of the global averaged response.

\subsection{Response dependence on the asymmetrical part of $\bf\Delta$}
\begin{figure}
    \includegraphics[scale=0.45]{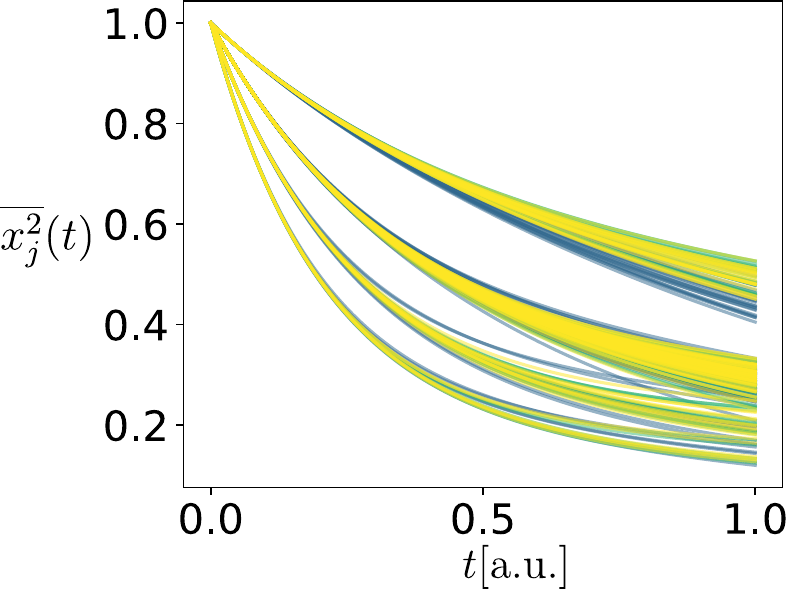}
    \caption{Square of the averaged response at each node of the network for three different structural perturbations (yellow, blue and green solid lines). 
    The matrix $\bf A$ is the same as in Fig.~\ref{fig1}(c) in the main text.
    The three structural perturbations are randomly drawn similarly as in Fig.~\ref{fig1} in the main text and Fig.~\ref{figS1}\,, but making sure that they all have the same symmetrical part $({\bf\Delta}+{\bf\Delta}^\top)$\,.
    }\label{figS2}
\end{figure}

In Fig.~\ref{figS2}, we compare the local averaged response of one network for three different structural perturbations that have the same symmetric part, but distinct asymmetric ones. 
The three different structural perturbations are randomly drawn in a similar fashion as Fig.~\ref{fig1} in the main text and Fig.~\ref{figS1}\,. 
One observes that at very short times, the three different local averaged responses almost perfectly superimpose, and only start to differ at later times. 
This is predicted by Eq.~(\ref{eq:lo_long}) in the main text.


\begin{thebibliography}{27}
\bibitem [{rem()}]{rem1}%
  \BibitemOpen
\bibfield  {journal} {  }\href@noop {} {\bibinfo  {journal} {One can easily check that up to the fourth order in $t$\,, no contribution comes from the higher order terms in Eq.~(\ref{eq:res2})}\ }\BibitemShut {NoStop}%
\end{thebibliography}
\end{document}